\documentclass[11pt,a4paper]{article}

\usepackage{amsmath}
\usepackage{graphicx}
\usepackage[square,comma,sort&compress,numbers]{natbib}

\begin{document}

\title{Efficient sampling of atomic configurational spaces}
\date{}

\author{L\'\i via B. P\'artay$^1$, Albert P. Bart\'ok$^2$ and G\'abor Cs\'anyi$^3$} 

\maketitle
\begin {center}
{\scriptsize{$^1$University Chemical Laboratory, University of Cambridge,
Lensfield Road, CB2 1EW Cambridge, United Kingdom, $^2$Cavendish
Laboratory, University of Cambridge, J J Thomson Avenue, CB3 0HE Cambridge, United Kingdom
$^3$Engineering Laboratory, University of Cambridge, Trumpington Street, CB2 1PZ Cambridge, United Kingdom}}
\end {center}

\begin{abstract} 
We describe a method to explore the configurational phase space of chemical systems. It is based on the nested sampling algorithm recently proposed by Skilling~\cite{bib:skilling,bib:skilling2} and allows us to explore the {\em entire} potential energy surface (PES)
efficiently in an unbiased way. The algorithm has two parameters which directly control the trade-off between the resolution with
which the space is explored and the computational cost.  We demonstrate the use of nested sampling on Lennard-Jones (LJ) clusters. Nested sampling provides a straightforward approximation for the partition function, thus evaluating expectation values of arbitrary smooth operators at arbitrary temperatures becomes a simple post-processing step. Access to absolute free energies allows us to determine the temperature-density phase diagram for LJ cluster stability. Even for relatively small clusters, the efficiency gain over parallel tempering in calculating the heat capacity is an order of magnitude or more. Furthermore, by analysing the topology of the resulting samples we are able to visualise the PES in a new and illuminating way. We identify a discretely valued order parameter with basins and supra-basins of the PES allowing a straightforward and unambiguous definition of macroscopic states of an atomistic system and the evaluation of the associated free energies.
\end{abstract}


\section{Introduction}
The study of potential energy hypersurfaces (PES) by computational
tools is one of the most rapidly developing areas within chemistry and
condensed matter physics. The potential energy (or Born--Oppenheimer)
surface describes the energy of a group of atoms or molecules in terms
of the geometrical structure (the set of nuclear coordinates), with the
electrons in their ground state\cite{bib:energy_landscapes}.  The
local minima and the associated ``basins'' of the potential energy surface represent metastable states and
the global minimum corresponds to the stable equilibrium configuration
at zero temperature.  The saddle points (of index one) correspond to
transition states that link neighbouring local minima and, within the approximation of transition state theory, dominate the
processes that involve structural change in the atomic configuration.
The dimensionality of the PES scales linearly with the number of
atoms, however, the number of local minima is commonly thought to scale
exponentially\cite{bib:exp_loc_min}, which makes exploration of the
PES computationally very demanding. For soft matter, liquid and
disordered systems, the physics is often dominated by entropic
effects, and the calculation of free energies requires a sampling over
large regions of the PES. For solid state systems, the unexpected
discoveries of new low energy configurations in hitherto unexplored
parts of the configurational phase space have consistently appeared
prominently in leading scientific
journals\cite{bib:Si_exchange_mechanism, bib:Al_adatom,
  bib:highpressure_co2, bib:binaryLJ_Wales, bib:fourfold_Goedecker,
  bib:silan_Pickard, bib:ammonia_Pickard}.

The last decade has seen huge activity in designing simulation schemes
that map out complex energy
landscapes\cite{bib:liuvoth,bib:walesbogdan}.  Several methods have
been developed to map different kinds of energy landscapes, optimised
to discover different parts of the PES, applicable to different sorts
of problems.  Global optimisation methods include Basin
Hopping\cite{bib:wales_basin_LJ}, Genetic Algorithms
(GA)\cite{bib:ratashvartsburg,bib:abrahamprobert} and Minima
Hopping\cite{bib:goedecker}. Temperature Accelerated
Dynamics\cite{bib:montalentivoter} samples rare events while Parallel
Tempering\cite{bib:partemp_swendsen, bib:partemp}, Wang-Landau
Sampling\cite{bib:wang_landau} and Metadynamics\cite{bib:michelettilaioparrinello} 
enable the evaluation of relative free energies. 
Each method has its particular set of advantages and
disadvantages, but what they have in common is that they (except for
some implementations of GA) are, in practice,  all ``bottom up'' approaches, i.e. start
from known energy minima and explore neighbouring basins. The
essential difference between the methods is in how they move from one
basin to another.

A new sampling scheme, nested sampling, was recently introduced by
Skilling\cite{bib:skilling,bib:skilling2} in the field of applied probability and
inference, to sample probability densities in high dimensional spaces
where the regions contributing most of the probability-mass are
exponentially localised.  The data analysis method based on the same
sampling scheme has already found use in the field
of astrophysics\cite{multinest1}. In this paper we adapt the nested sampling approach for exploring
atomic configurational phase spaces and not only provide a new
framework for efficiently computing thermodynamic observables, but
show a new way of visualising the pertinent features of complex
energy landscapes

\begin{figure}[btp]
\begin{center}
\includegraphics[width=8cm]{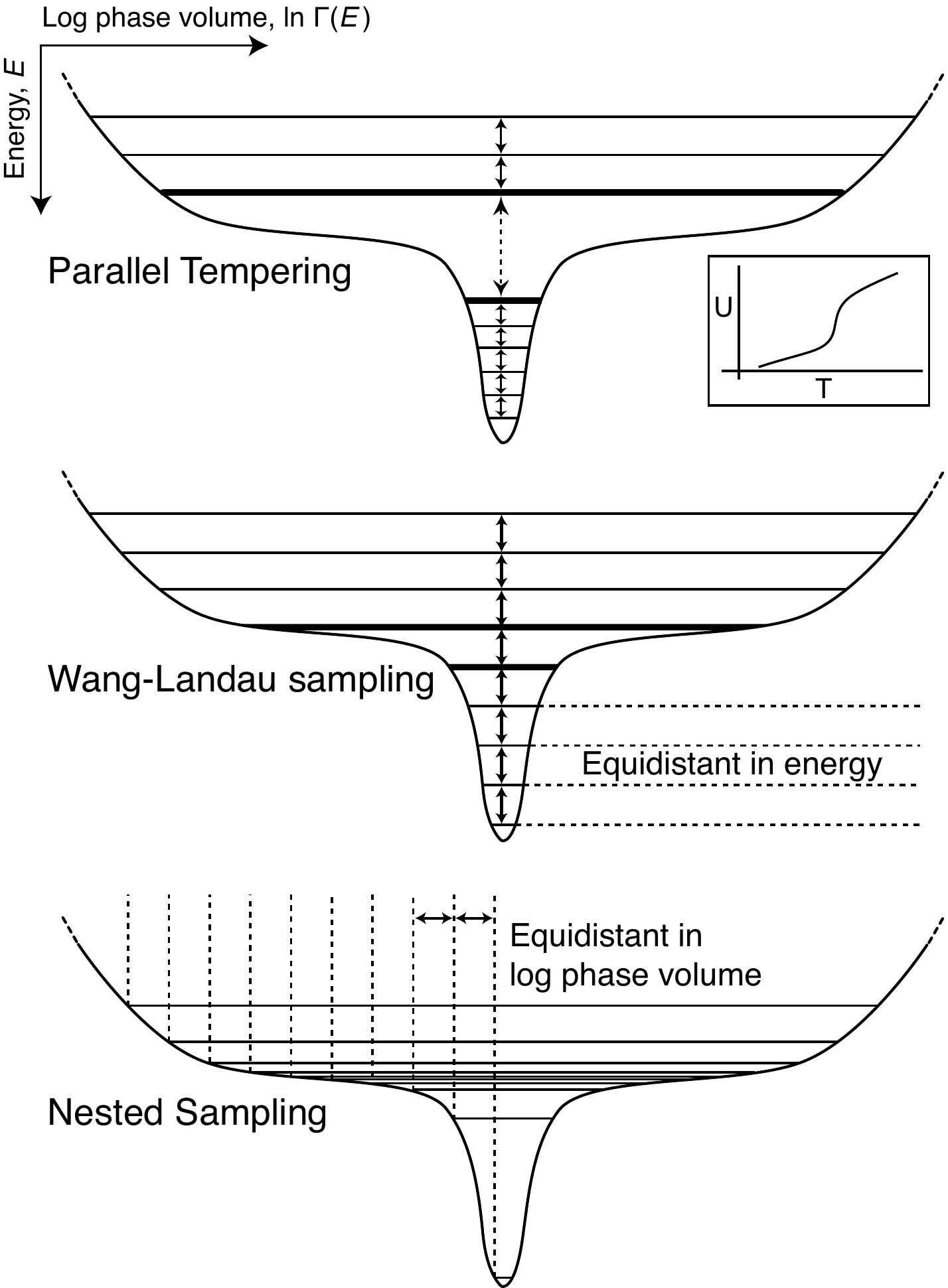}
\end{center}
\caption {Cartoon illustration of how parallel tempering (top panel), Wang-Landau sampling (middle panel) and nested sampling (bottom panel) deal with a first order phase transition.  In each panel, the width of the black curve at a given height represents the logarithm of the available configurational phase space, $\ln \left[\Gamma(E)\right]$, as a function of potential energy, $E$. This is an entropy-like quantity, but considered as a function of potential energy, rather than temperature. The series of sampling regions (in temperatures or energies) are represented by horizontal lines. The thick lines on the top two panels on either side of the phase transition show the entropy jump that makes equilibration difficult. Nested sampling avoids this by constructing a series of energy levels equispaced in $\ln \Gamma$.}
\label{fig:sampling_methods}
\end{figure}

To demonstrate the key idea of nested sampling, we illustrate its behaviour in Figure~\ref{fig:sampling_methods} for a first order phase transition,
and compare it with parallel tempering and Wang-Landau sampling. 
A first order phase transition is characterised by a dramatic reduction of the available phase space as the 
energy of the system is reduced. All three methods operate by simultaneously or sequentially sampling states of the system at a 
series of energy or temperature levels (for a large system, these are equivalent). For the overall sampling to be successful, the 
samplers at each level have to equilibrate with those on every other level. Let us consider parallel tempering first, which samples at 
fixed temperatures (top panel). The overlap between the distributions at levels just above and just below the phase transition is very 
small, vanishing in the thermodynamic limit, due to the entropy jump. It is well understood that this makes equilibration of the 
samplers in the two phases especially difficult\cite{marinari_chapter1996}.  Furthermore, due to the steep increase of the energy 
near the transition temperature (see inset), the spacing of the corresponding energy levels becomes wider near the phase transition, 
making equilibration even harder. 

Wang-Landau sampling (middle panel) is done on energy levels constructed to be equispaced, but 
large systems are still very difficult to equilibrate due to the entropy jump, resulting in broken ergodicity. This phase equilibration 
problem can be solved by constructing a sequence of energy levels in such a way that the phase space volume ratios corresponding 
to successive energy levels is an $O(1)$ constant, as shown on the bottom panel. Hence the energy level spacings near 
the phase transition will be narrow, precisely what is needed to allow good equilibration. The nested sampling algorithm, described below, constructs just such a sequence of energy levels using a single top-down pass. This sequence is similar in spirit to the sequence of temperatures that would be obtained at {\em end} of a parallel tempering run which adapts the temperature levels until all exchanges between neighbouring levels is the same $O(1)$ constant\cite{adaptive_partemp1,adaptive_partemp2}.

\section{Nested sampling}

The fundamental principle of statistical mechanics is that, in equilibrium, all accessible states of an isolated system occur with equal probability. Accordingly, the expectation value of an observable $A$ is the simple sum over the microstates of the system,
\begin{equation}
  \langle A\rangle =  \frac{1}{Z} \sum_{\{x,p\}} A(x, p),
\label{eq:obs1}
\end{equation}
where $x$ and $p$ are the positions and momenta, respectively, and the partition function $Z$ is the normalization constant, in this case just the total phase space volume. However, when a system is in thermal equilibrium with its surroundings, the occupation probability of its internal states are given by the canonical distribution and the expectation value of the observable is
\begin{equation}\label{eq:obs}
  \langle A\rangle =  \frac{1}{Z(\beta)} \sum_{\{x,p\}} A(x, p) e^{ -\beta H(x,p)},
  \end{equation}
where $H$ is the Hamiltonian of the system, $\beta$ is the inverse thermodynamic temperature and $Z(\beta)$ is the canonical
partition function, 
\begin{equation}\label{eq:Zthermo}
Z(\beta) = \sum_{\{x,p\}} e^{-\beta H(x,p)} 
= Z_p(\beta) \left(\frac{\delta V}{V^N}\right) \sum_{\{x\}}  e^{-\beta E(x)},
\end{equation}
where $E$ is the potential energy, $V$ is the volume, $\delta V$ is the volume element of the spatial discretisation,  $N$ is the number of particles and the momentum-dependent part has been separated out as usual,
\begin{equation}\label{eq:Zp}
Z_p(\beta) = \left( \frac{2 \pi m}{\beta h^2} \right)^{3N/2} \frac{V^N}{N!},
\end{equation}
where $m$ is the mass of the particles and $h$ is Planck's constant. Defining $w \equiv \delta V/V^N$, the position-dependent part of the partition function is
\begin{equation}
Z_x(\beta) = \sum_{x} w e^{-\beta E(x)}
\end{equation}
with
\begin{equation}
\sum_{\{x\}} w = 1.
\end{equation}

Let us assume that the operator $A$ does not depend explicitly on the momentum variables, and consider
estimating $A$ and $Z_x(\beta)$ directly by turning them into sums over a set of
{\em sample points} $\{x_i\}$, 
\begin{eqnarray}
  \langle A\rangle_\mathrm{est} &=&  \frac{1}{Z_\textrm{est}(\beta)} \sum_i w_i A(x_i) 
e^{-\beta E(x_i )}\label{eq:obssample}\\
Z_\textrm{est}(\beta) &=& \sum_{i} w_i  e^{-\beta E(x_i)}\label{eq:zsample},
\end{eqnarray}
where $w_i$ is the (now not necessarily uniform) weight factor representing the volume element associated with the sample point $x_i$, still with the normalisation
\begin{equation}
\sum_i w_i = 1.
\end{equation}
In general, it is difficult to find a set of sample points that gives a good coverage of phase space and 
a good approximation of observables and the partition function at low and moderate temperatures because the Boltzmann factor is very sharply peaked near
states with low energy, and such states occupy an extremely small volume in phase space. This, in essence, is the ``sampling problem'' in molecular simulation. Note that in principle, equations~\ref{eq:obssample} and \ref{eq:zsample} could be used to estimate the partition function using samples from a regular Monte Carlo simulation by evaluating the ensemble average of the observable $A=\exp(\beta E(x))$,   
\begin{equation}\label{eq:average_boltz}
\langle e^{\beta E(x)} \rangle_{NVT} = \frac {1}{Z_{x}} \sum_i w_i e^{\beta E(x_i)} e^{-\beta E(x_i)}  = \frac {1} {Z_x}.
\end{equation}
However, in practice, the above sum is essentially impossible to evaluate in finite time because the variance of $\exp(\beta E)$ is divergent at low temperatures. This approximation is called the "harmonic mean approximation" in the probability literature\cite{bib:newton-raftery}, for further details see the discussion by C.~P.~Robert and N.~Chopin\cite{bib:disscussion_Robert}.  

\begin{figure}[btp]
\begin{center}
\includegraphics[width=8cm]{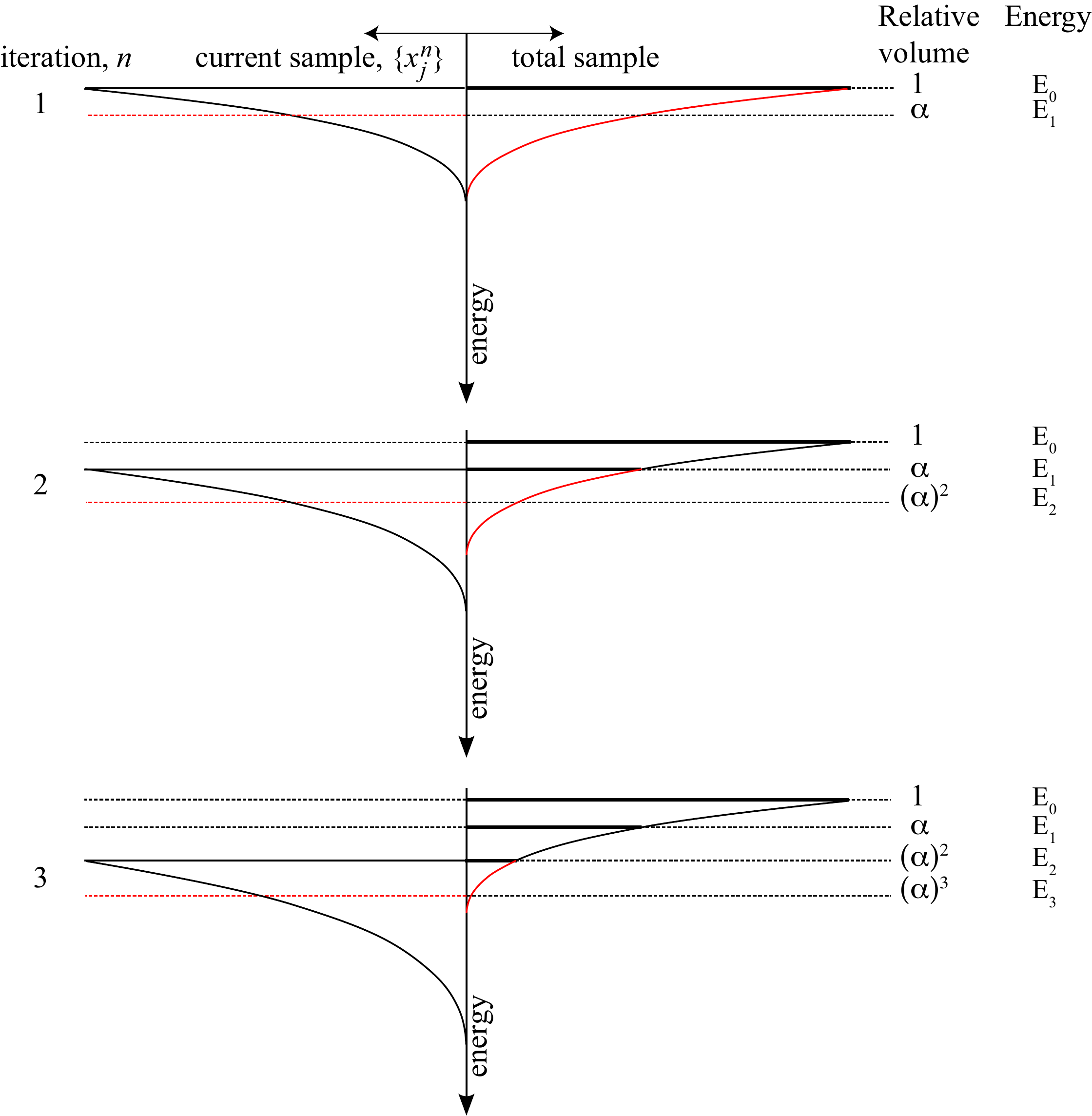}
\end{center}
\caption {Illustration of the nested sampling iterations. Each curve represents an energy histogram of samples, curves on the left are the samples of the current iteration, while those on the right are the weighted histograms of all the samples collected. The red part of the total histograms on the right represents the new contribution, appropriately scaled, from the current iteration on the left.  After every iteration, once the new samples have been collected, the new lowest energy level is defined to be at a fixed percentile (as a fraction $\alpha$) of the current sample histogram, illustrated by the red dashed line. This ensures that the new energy level corresponds to a phase space volume which is a fraction $\alpha$ smaller than the previous one.}
\label{fig:ns_method}
\end{figure}

In the nested sampling method, illustrated in Figure~\ref{fig:ns_method}, we instead break up the sums in \ref{eq:obssample} and \ref{eq:zsample} into terms associated with a series of decreasing energy levels $\{E_n\}$ each with a corresponding weight factor, which are also decreasing. For each level, a set of $K$ sample points $\{x^n_j\}$ is obtained by uniform sampling from the energy landscape {\em below} $E_n$:
\begin{equation}\label{eq:ns_samples}
x^n_j \sim U(\{x: E(x) < E_n\}). 
\end{equation}
After iteration $n$ we create the new lowest energy level $E_{n+1}$ at a fixed percentile, denoted by fraction $\alpha$, of the energy distribution of the samples at level $n$. Thus, the sample points corresponding to each energy level will have a combined phase space volume that is a factor $\alpha$ smaller than that of the previous level, so the phase space volume of the $n$th level is  $(\alpha)^n$. The sample points that lie between levels $E_n$ and $E_{n+1}$ contribute to the overall sample with a weight $w_n$ that is the difference between the phase space volumes corresponding to the two levels, $w_n = (\alpha)^n-(\alpha)^{n+1}$.  Since the phase space volumes decrease exponentially for a fixed $\alpha$, the overall sampling converges quickly, and is able to locate exponentially small parts of configurational space. At the same time, because every pair of neighbouring energy levels has a phase space volume ratio of $\alpha$ (even near energies that correspond to discontinuous phase transitions), there is no inherent difficulty in generating the samples, there is no "equilibration problem", see bottom panel of Figure~\ref{fig:sampling_methods}.

After a sufficient number of iterations, using all the samples from eq~\ref{eq:ns_samples} for all levels, the nested sampling approximation of the configurational partition function and of an observable becomes
\begin{eqnarray}\label{eq:ns_op} 
Z_\textrm{est}  &=& \sum_n \left[(\alpha)^n-(\alpha)^{n+1}\right] \sum_{j: E_n < E(x_j) < E_{n+1}} e^{-\beta E(x_j)}\\ 
\langle A\rangle_\textrm{est} &=& \frac{1}{Z_\textrm{est}} \sum_n  \left[(\alpha)^n-(\alpha)^{n+1}\right] \sum_{j: E_n < E(x_j) < E_{n+1}} A(x_j) e^{-\beta E(x_j)},
\end{eqnarray}
where $n$ runs over the energy levels and for each level $j$ runs over those sample points that lie between the successive pair of levels.
To complete the description of the algorithm, it remains to specify how the samples $x^n_j$ are drawn 
in eq~\ref{eq:ns_samples}. We use a simple rejection Gibbs sampler\cite{bib:JNeumann,bib:mackaybook}, in the form of a strictly bounded random walker that at each level $n$ is constrained to remain inside the region $\{x: E(x) < E_n\}$. The nested sampling algorithm as described here is the generalisation of  the work of Skilling and equivalent to it for the choice $\alpha=(K-1)/K$. For this special choice, which we also adopt in the rest of this work, the sum over $j$ in eq~\ref{eq:ns_op} has only one term, since precisely one sample is contained in the top $1/K$ fraction of each sample set. 

Thermodynamic quantities are obtained from the partition function, for example, the heat capacity is given by
\begin{equation}\label{eq:heat_capacity}
C_V = \left ( \frac {\partial U} {\partial T} \right)_V = -
      \left ( \frac {\partial}{\partial T} \frac {\partial \ln Z} {\partial \beta} \right)_V.
\end{equation}
The expectation value of the internal energy can be written, in terms of the samples, as
\begin{equation}\label{eq:energy}
   U = -\frac {\partial \ln Z} {\partial \beta} \approx \frac{3N}{2}
   \frac {1} {\beta} + \frac {1} {Z_\textrm{est}}
   \sum_{n} w_n E_n \exp(-\beta E_n) 
\end{equation}
and its derivative with respect to the temperature as
\begin{equation}\label{eq:ns_heat_capacity}
\begin{split}
\left ( -\frac {\partial} {\partial T}
\right)_V \frac {\partial \ln Z} {\partial \beta} \approx \frac {3N} {2} k - 
       & \frac {\sum_{n} w_n E_n\exp(-\beta E_n)/kT^2}
              {Z_\textrm{est}^2} 
        \sum_{n} w_n E_n\exp(-\beta E_n)  + \, \\
       & \frac {1} {Z_\textrm{est}}
        \sum_{n} w_n E_n^2 \exp(-\beta E_n)/kT^2.
\end{split}
\end{equation}

The convergence of the nested sampling procedure has been extensively discussed\cite{bib:skilling_convergence,bib:Evans}, and the error in $\ln Z$ scales as $K^{-1/2}$. Note that in contrast to the case of general distributions, the convergence results in statistical mechanics are easier to obtain because the energy is bounded from below. 

Finally, note the absence of the temperature $\beta$ in the actual sampling
algorithm. Since $\exp(-\beta E)$ is a monotonic
function of $E$, the above derivation of the sampling weights is
independent of $\beta$. Thus the expectation value of any observable
can be evaluated at an arbitrary temperature just by resumming over
the same sample set, obviating the need to generate a new sample set
specific to each desired temperature.  The exponential
refinement of the sampling for low energies becomes increasingly less
relevant (but not incorrect) at higher temperatures for which the low
energy states contribute less to the partition function. This athermal
aspect of the sampling scheme is similar to that of the Wang--Landau
method\cite{bib:wang_landau, bib:WL_LJ_liquidvapour,
  bib:wales_WL_BS}. However, the convergence
problems\cite{bib:WL_disadvantages, bib:WL_disadvantages2} that
typically arise for systems with broken ergodicity are not present in
our case, due to the top-down nature of the method: the samples are uniformly distributed at each energy level, and the low energy samples are
directly obtained from the higher energy ones. For a given $K$, nested sampling always converges, and $K$ determines
the resolution with which we sample the basins of the PES. If a
particular basin in its energy range has a phase space volume ratio relative to
the rest of the space that is smaller than about $1/K$, the
probability that a sample point will ever fall into that basin is
small. Therefore, by increasing $K$, we are able to explore the PES
with higher resolution. Notice how this limited resolution is
related to an effective minimum temperature: if a sampling set explores
basins whose phase space volumes are typically larger than some value,
then there will be a corresponding temperature above which these basins will dominate
the behaviour of the system due to their entropy.


\section{Lennard-Jones clusters}

We demonstrate the new framework in the
context of Lennard-Jones (LJ) clusters, which is a favourite testing ground
for new phase space exploration schemes, partly because the potential
energy function is cheap to calculate and partly because an enormous
amount of data has been amassed about the potential energy
landscape in the literature\cite{bib:energy_landscapes,bib:wales_basin_LJ,bib:wales_globopt_LJ,
bib:vladimir}.  

\begin{figure}[btp]
\begin{center}
\includegraphics[width=10cm]{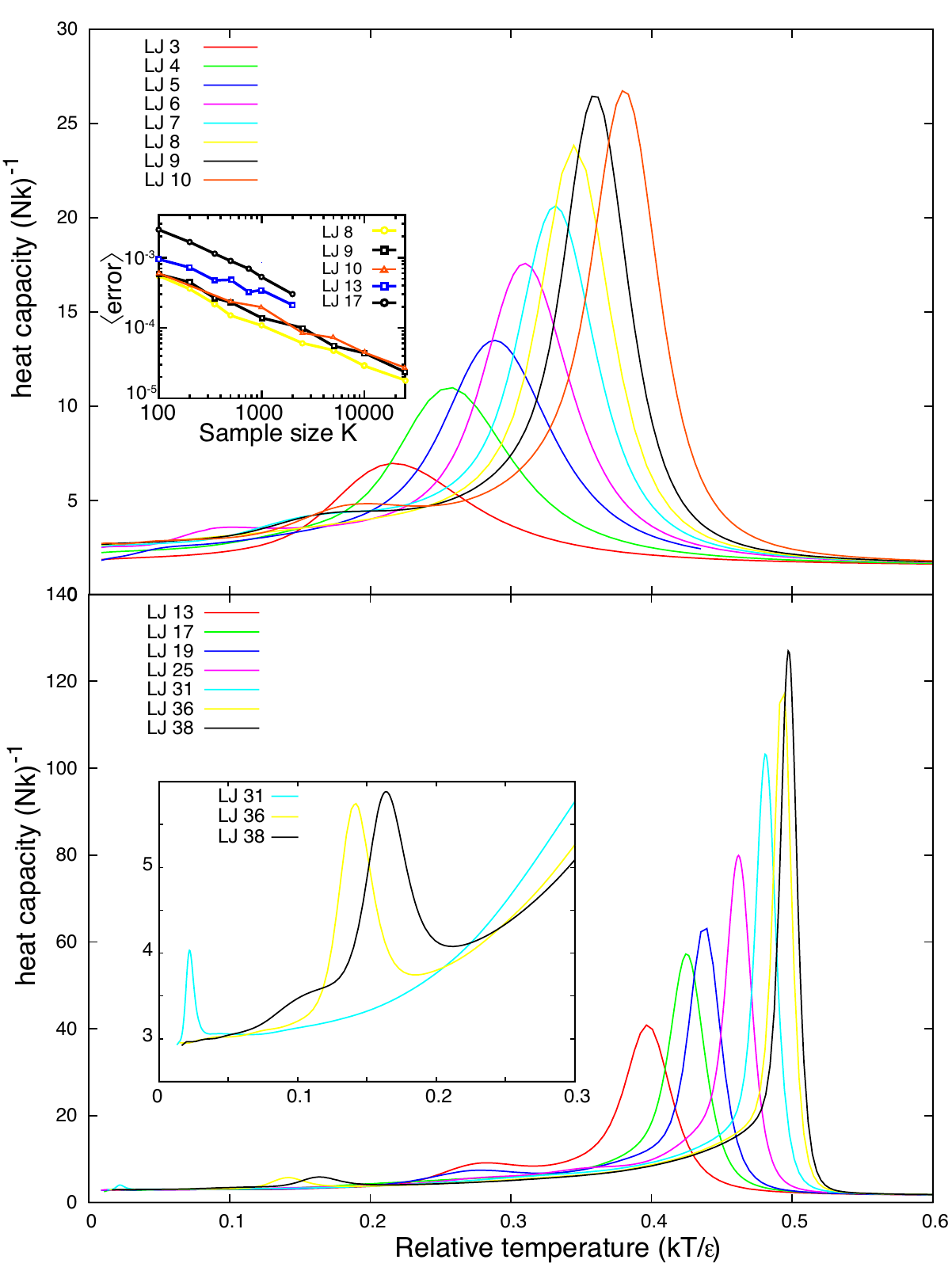}
\end{center}
\caption {Heat capacity as a function of temperature, for
  Lennard-Jones clusters containing less (top) and more (bottom) than
  10 atoms. The inset of the top panel shows the average error of 
  the heat capacity as a function of sample size (logarithmic scale), computed relative to the largest sample.
  The inset of the bottom panel shows the magnified low temperature region of the heat capacity curves of larger clusters.}
\label{fig:LJ_heat_capacity}
\end{figure}

We performed nested sampling on a number of LJ particles, with potential parameters $\varepsilon$ and $\sigma$, in a periodic box, corresponding to a low
density of $2.31 \times 10^{-3} \sigma^{-3}$, using a cutoff of 3$\sigma$,
such that at low temperature the particles aggregate into a
cluster.  The heat capacities of small LJ clusters are shown in
Figure~\ref{fig:LJ_heat_capacity}, calculated according to eq~\ref{eq:ns_heat_capacity}.  
Note that since the nested sampling procedure is independent of temperature,
only one simulation was needed for each cluster size, after which the heat capacity is trivial to 
evaluate at an arbitrary temperature.  For each cluster, a shoulder and a large
peak is present, corresponding to melting and sublimation. For our largest clusters with 31, 36 and 38 atoms, the new
peak at low temperature, discussed in more detail below, is traditionally associated with the so-called Mackay--anti-Mackay transition\cite{bib:M-aM,bib:M-aM_Northby}.
  
 The size of the sample set was
increased until convergence of the heat capacity was achieved, and it is shown in Table~\ref{table:convergence}, together with the number of energy evaluations performed during the calculations. The inset of Figure~\ref{fig:LJ_heat_capacity} shows
the average error in the heat capacity as a function of the sample size. The convergence
shows a power law behaviour with a slope that corresponds to an $O(K^{-1/2})$ error, typical in statistical sampling. 
  
\begin{figure}[btp]
\begin{center}
\includegraphics[width=7cm]{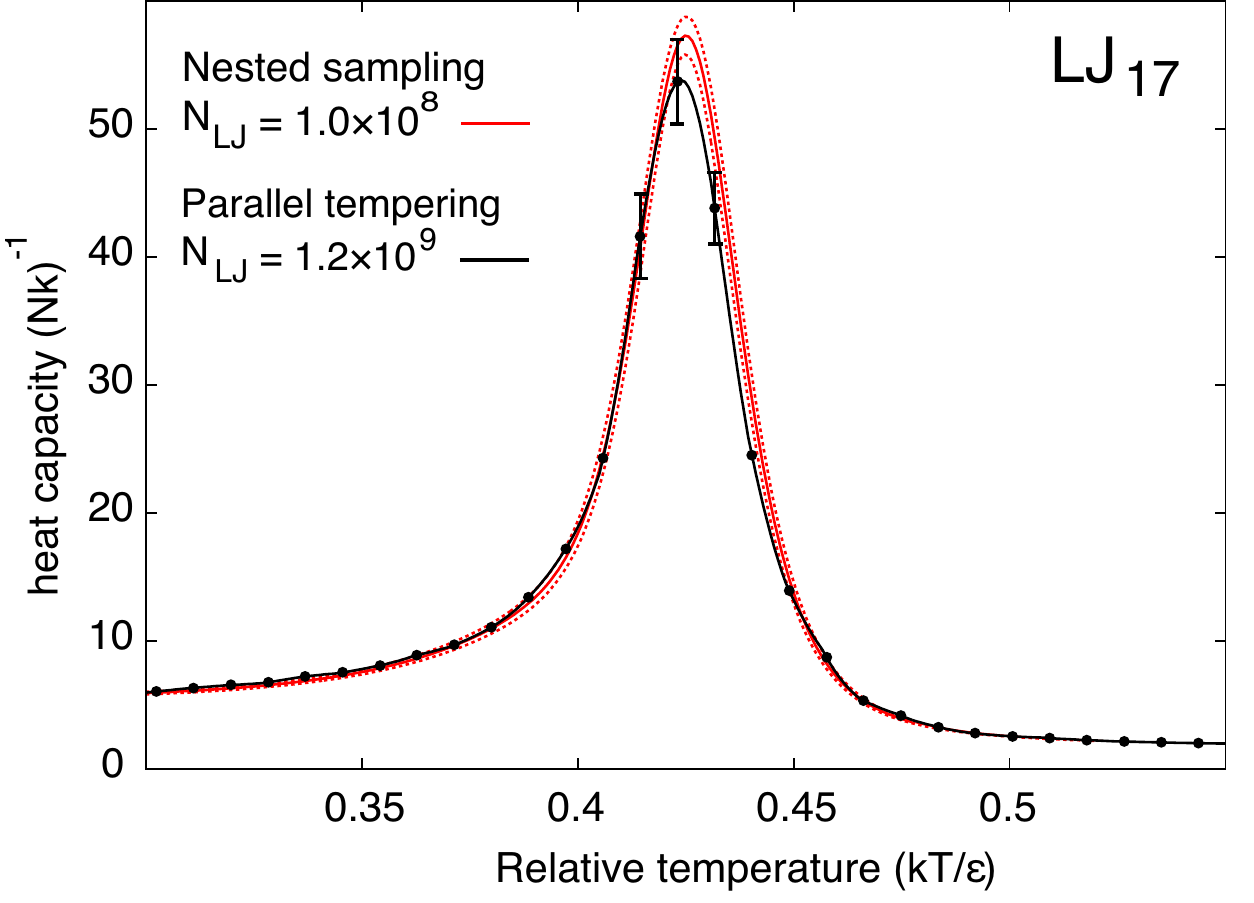}
\includegraphics[width=7cm]{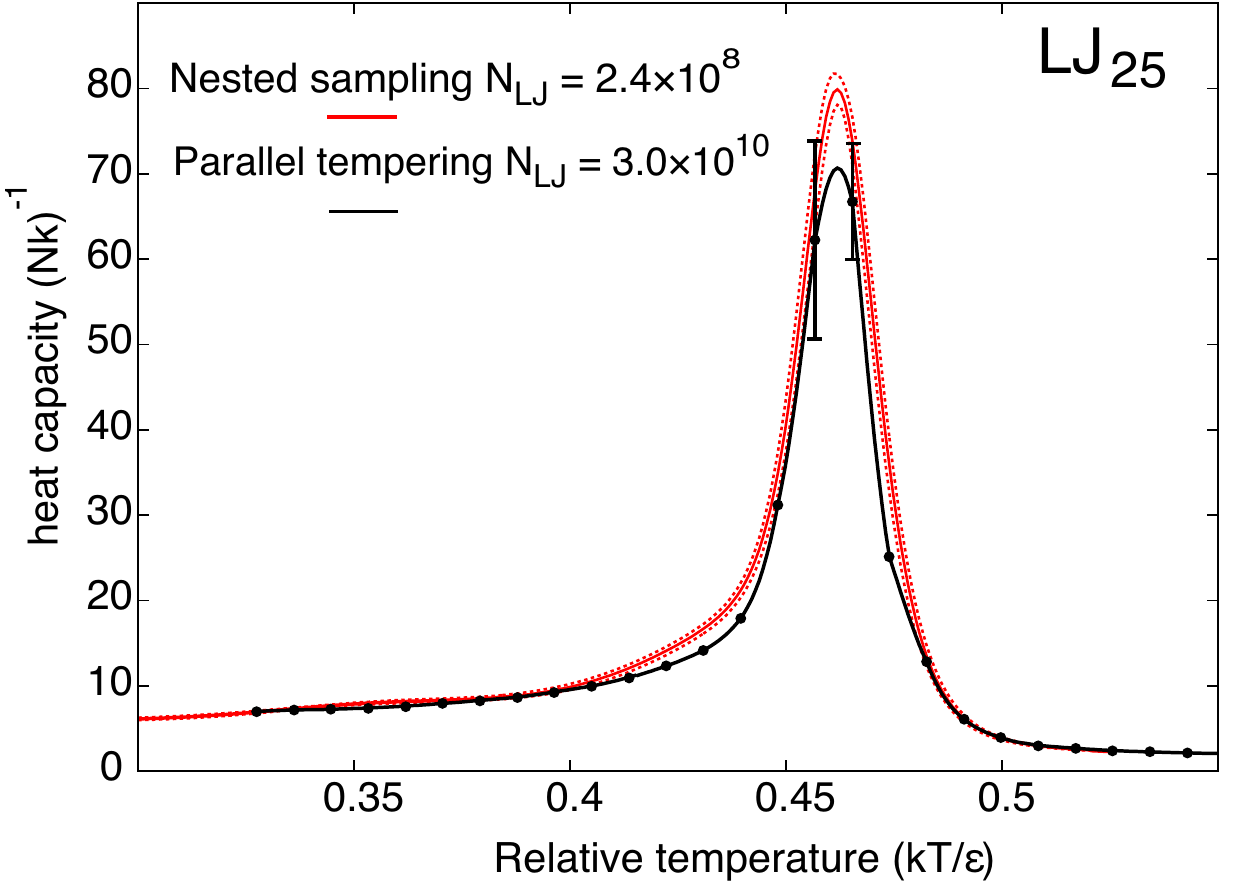}
\end{center}
\caption {Heat capacity as a function of temperature for
  Lennard-Jones clusters LJ$_{17}$ and LJ$_{25}$, using nested sampling (red lines) and parallel 
  tempering (black lines with black dots indicating the temperature values where the simulations were performed in parallel tempering). The standard errors of the prediction are shown by dashed lines (nested sampling) and error bars (parallel tempering). The number of energy evaluations needed to calculate the curves are also indicated.} 
\label{fig:LJclusters_NSvsPT}
\end{figure}

In order to demonstrate the power of nested sampling, we chose 
two Lennard-Jones clusters and calculated the heat-capacity 
curves using parallel tempering.
The resulting comparison is shown in Figure~\ref{fig:LJclusters_NSvsPT}. 
For the parallel tempering calculations we chose a set of equispaced temperatures with a spacing that can resolve the peaks.  To achieve a fair comparison, both methods were converged until the respective maximum errors in the heat capacities were about the same. As mentioned above, nested sampling automatically yields the observables as continuous function of temperature. In the case of parallel tempering, where heat capacity values are observed directly only at the discrete set of temperatures, to obtain values at intermediate temperatures we used Boltzmann reweighting on the samples from the Markov chains at successive pairs of temperatures\cite{boltzmann-reweighting}.  For LJ$_{17}$ the efficiency gain of nested sampling is a factor of 10 over parallel tempering, while for the larger LJ$_{25}$ cluster, as the entropy jump is larger, the efficiency gain is a factor of 100. 

Our heat capacity curves for the largest clusters (LJ$_{31}$, LJ$_{36}$ and LJ$_{38}$) are consistent with what is reported in the literature \cite{bib:vladimir,LJ3138partemp}, using computational resources of similar order of magnitude. Note however, that advance knowledge of the global minimum is not required and was not used in the nested sampling simulations. The relatively large computational cost for LJ$_{31}$ and LJ$_{38}$ is precisely because these clusters have global minima which are hard to find\cite{wales_LJ_disconn}. See below for a more detailed discussion of the relative sizes of the basins containing the global and lowest local minima. 

\begin{table}
\caption{Size of sample set and number of energy evaluations needed to produce a converged 
nested sampling run (so that the error in the heat capacity curve is on the order of the line thickness of Figure~\ref{fig:LJ_heat_capacity}).}
\begin{tabular}{crc}
& \\ [-1.5ex]
\hline\hline
LJ cluster size& Size of sample set, $K$ & Number of energy evaluations  \\
\hline
 2-5    & 300 &$1.6 \times 10^6$ \\  %
 6-10   & 500 &$2.0 \times 10^7$ \\  %
11-15   & 700&$1.1 \times 10^8$ \\  %
16-20   & 900 &$2.2 \times 10^8$ \\  %
21-25   & 1\,100 &$3.9 \times 10^8$ \\  %
31   & 288\,000&$3.4 \times 10^{12}$ \\
36   & 32\,000 &$2.8 \times 10^{11}$ \\
38   & 224\,000&$2.6 \times 10^{12}$ \\
\hline\hline
\end{tabular}
\label{table:convergence}
\end{table}


\section{LJ cluster phase diagram}

\begin{figure}[bthp]
\begin{center}
\includegraphics[width=10cm]{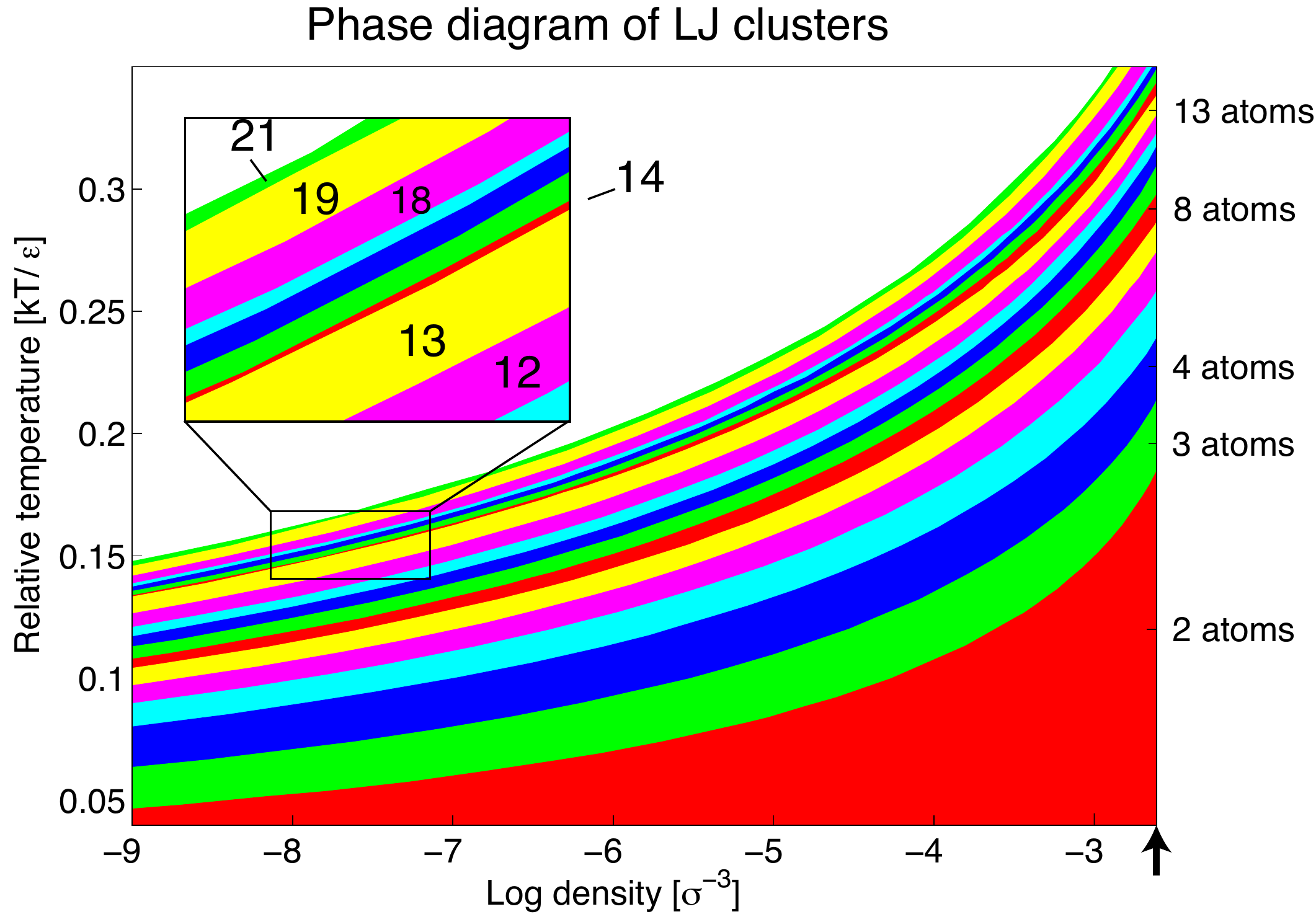}
\end{center}
\caption {Phase diagram of Lennard-Jones clusters as a function of temperature
  and density. Each coloured band represents a region in which the
  corresponding cluster is thermodynamically stable against
  evaporation while the smaller clusters are not. The black arrow indicates the density at which the nested sampling calculations were carried out.}
\label{fig:cluster_stability}
\end{figure}

The ability to compute the partition function and hence the absolute
free energy $A$, as 
\begin{equation}
A(\beta) = - \frac{1}{\beta} \ln Z(\beta)
\end{equation} 
enables us to plot in Figure~\ref{fig:cluster_stability} a phase diagram
showing the stability of the Lennard-Jones clusters against the ideal gas
(i.e. evaporation) as a function of density and temperature. The
total partition function of the ideal gas is $Z_p$ (see eq~\ref{eq:Zp})
since the potential energy is zero. We 
performed the nested sampling calculations at a single density (marked by an arrow on Figure~\ref{fig:cluster_stability}), in which individual clusters are formed that do not interact with their periodic images. The partition function for lower densities can be approximated by multiplying the partition function by a $V'/V$ factor, where 
the volume $V'$ corresponds to the new density, thus the translational freedom of the cluster in a larger volume is included. Finally, for every Lennard-Jones cluster at a given density, the 
critical temperature at which its free energy 
becomes larger than that of the ideal gas at the same density can be determined.
In Figure~\ref{fig:cluster_stability} each coloured area represents 
a region inside which the corresponding cluster is stable, i.e. $A_{{\mathrm{LJ}}_N} < A_{\textrm{LJ}_{N-1}}+A_{\mathrm{LJ}_1}$. 
Larger clusters are more stable, thus the regions
form a nested sequence of bands that correspond to
areas where a given cluster is stable but smaller clusters are
not. Particularly favourable clusters show up in this
diagram.  The band corresponding to LJ$_{13}$ is wider than its
neighbouring bands, mostly obscuring the region corresponding to
LJ$_{14}$. LJ$_{19}$ is so much more favourable than LJ$_{20}$ that
there is no region where the latter is stable and the former is not. 
Note that this is a phase diagram for clusters only, it does not include cluster-cluster
interactions, and so does not extend to high densities where solid phases are formed. 

\section{Energy Landscape Charts} 

Visualisation of the energy landscape can greatly enhance the understanding of a chemical system, but the representation of a $3N$ dimensional function is a challenging task. 
One way to get around this problem is to reduce the dimensionality by projecting the energy landscape 
onto ad-hoc collective coordinates, but this does not provide a sufficient description in general, and it can be very misleading in some cases, depending on the choice of collective coordinates.
A more adequate and usual way of depicting the topology of basins and transition states is the disconnectivity
graph\cite{bib:disconnectivity, bib:wales_disconnectivity}, or the scaled disconnectivity graph\cite{bib:wales_scale_disconnectivity},
where in the latter case the width of the graph is made proportional to the number of minima. While the complete disconnectivity graph would capture the entire landscape, it is impractical to calculate or draw for even moderate size systems, and more importantly, it does not contain the entropy information which controls the thermodynamic behaviour. Nested sampling naturally provides a solution to this.

In this section we introduce and illustrate an algorithm that identifies the large scale basins of the energy landscape 
 automatically by post-processing the sample set produced by nested sampling. The key point is that we get a broad-brush view of the landscape, using relatively few samples (clearly not enough to discover all local minima), but nevertheless giving a helpful overview of the system. 
To carry out the topological analysis of the samples, we construct a
graph in which the vertices are the sample points, and connect them by edges based on
the Cartesian distance between the sample points: each vertex is
connected to its $k$ nearest neighbours which have a higher energy
than itself. Then we successively remove vertices and their associated
edges from the graph in a decreasing order in energy. When the removal of a vertex results in the
graph splitting into two or more disconnected subgraphs, the vertices in the subgraphs are identified with new basins. 
The relative phase space volumes of the basins
is estimated from the ratio of the number of samples belonging to each at the
moment of splitting. The subgraphs are analysed
recursively using the same procedure. If a subgraph is eliminated
without splitting further, it represents a basin associated with a
local minimum, and we identify the sample with the lowest energy in
this basin as our estimate of the local minimum. The output of the algorithm is a hierarchical nested tree of basins, with known phase space volumes. 

\begin{figure}[tbhp]
\begin{center}
\hbox{
\vbox{
\hbox{\hskip 0.6cm\includegraphics[width=7cm]{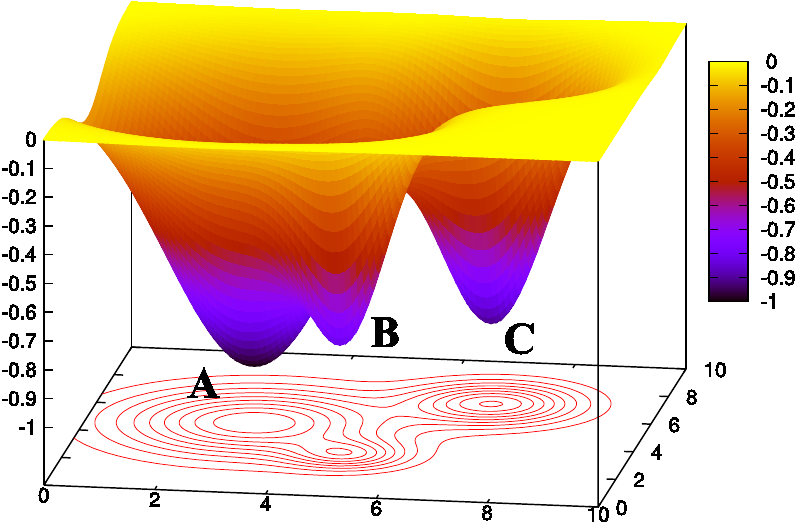}}
\hbox{\includegraphics[width=7cm]{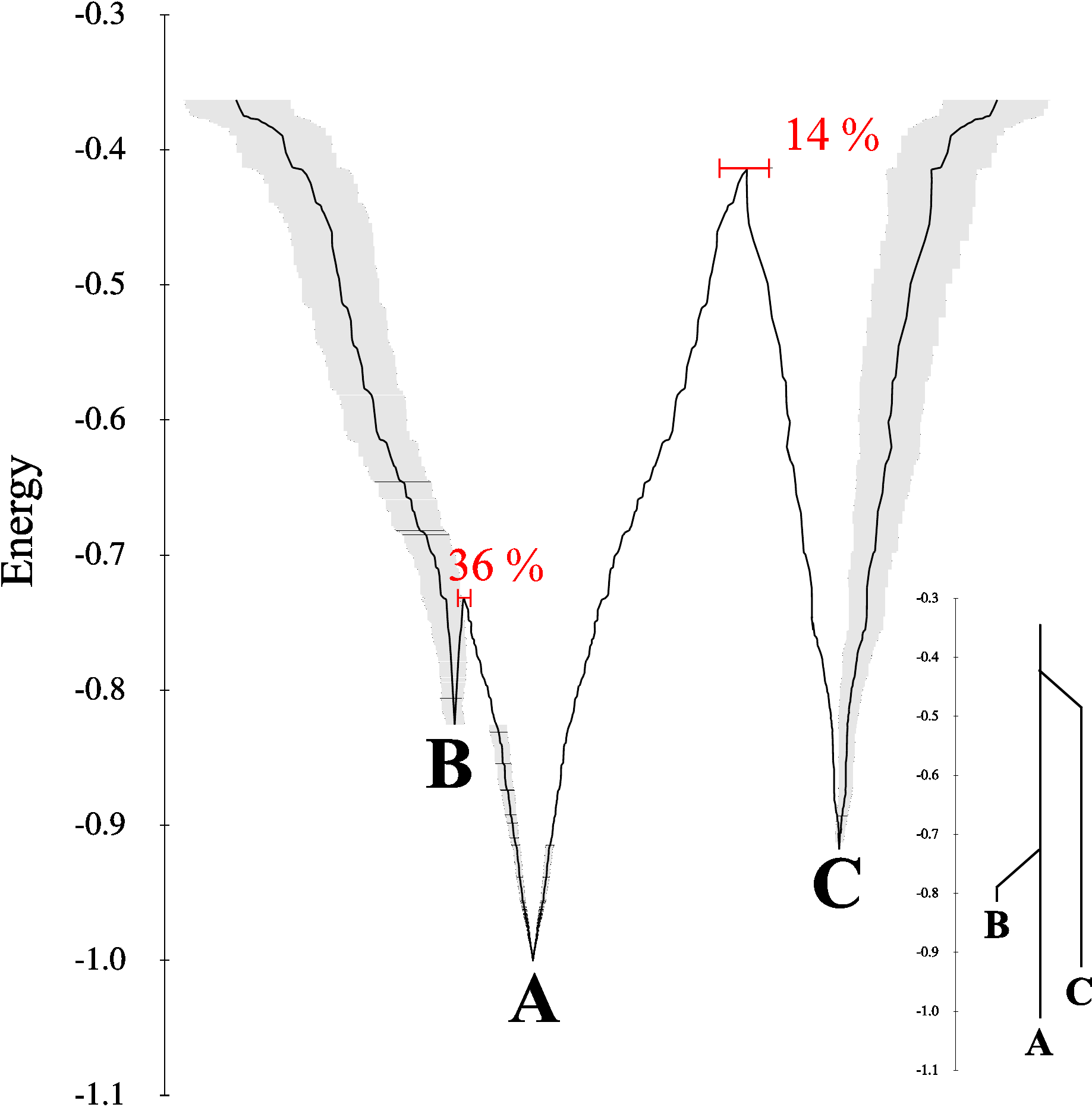}}
}
\includegraphics[width=8cm]{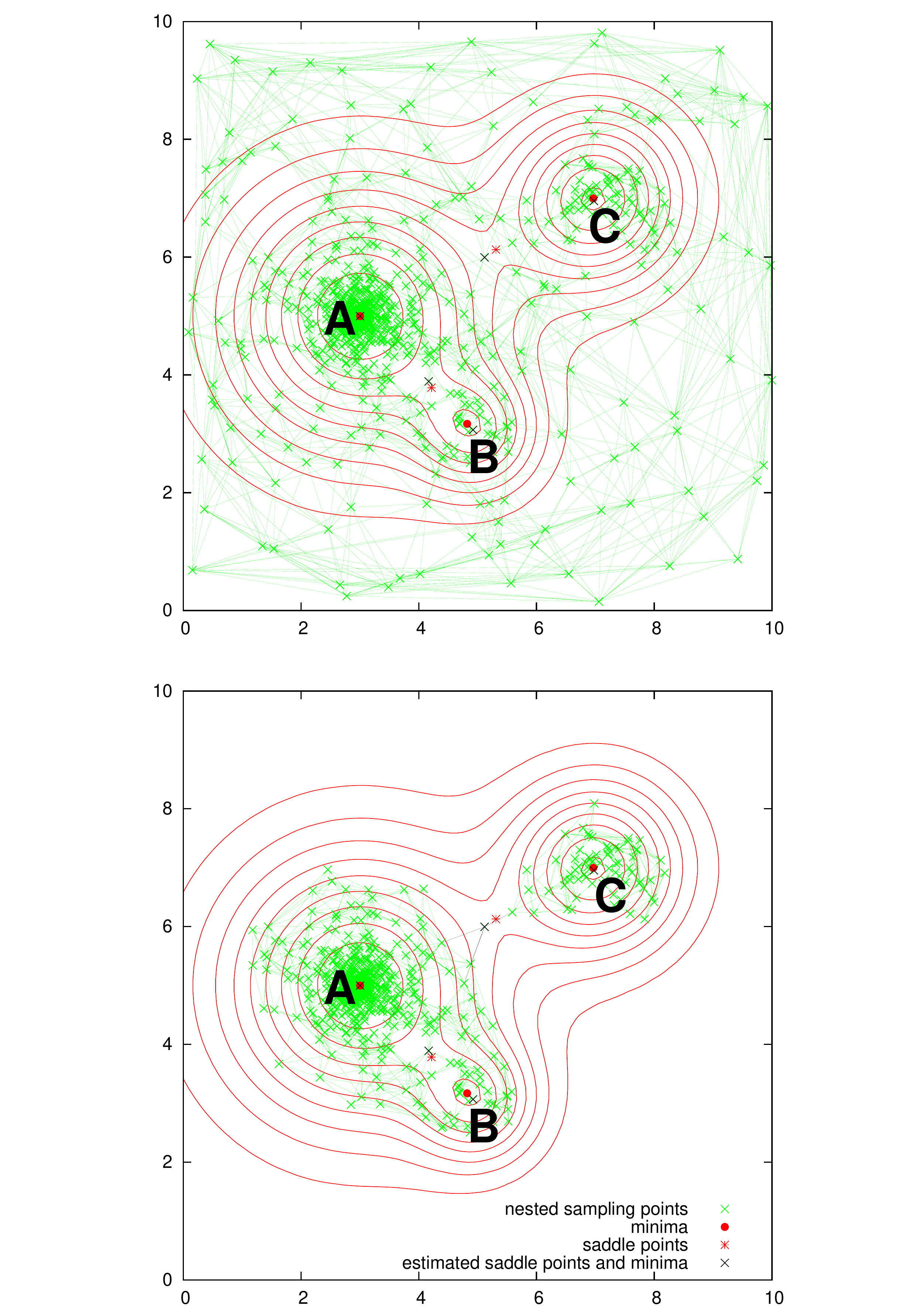}
}
\end{center}
\caption {Real energy landscape (top left panel) and the chart produced by nested
  sampling (bottom left panel) for the toy model, together with the corresponding disconnectivity graph. 
  The global minimum is marked by A, while the two local minima are marked by B and C. The vertical scale is the
  energy, the horizontal dimension on the landscape chart represents the phase space volume
  enclosed by the set of samples at a given energy, separated out into
  different basins. This is achieved using a geometric analysis of the sample set, as
  described in the text. 
  The
  gray shading represents the error in the overall phase space
  volumes, while the red lines indicate the error in the relative
  volumes of the three basins. The percentage figures refer to the
  relative size of the error as compared to the volume of the smaller
  of the basins at the energy level where the basins separate.
  The sample configurations and the graph constructed from them are shown on the right panels. The real minima and transition
  states are shown by red dots and stars, respectively, as well as the
  corresponding estimates from post processing the nested sampling data
  (see text). Top right : full graph; bottom right: in the process of elimination
  of vertices in order of decreasing energy, the step in
  which the graph is about to split into two identifies the sample
  point close to the saddle point.}
\label{fig:test_3min_3D}
\end{figure}

To demonstrate this procedure, we show
how it works on a simple toy model, a two dimensional potential energy
surface given by the sum of three Gaussians, shown in the top left panel of
Figure~\ref{fig:test_3min_3D}. This surface has two local minima
in addition to the global minimum.
We performed a nested sampling run on this surface using $K=100$
samples and 1900 iterations, in this case choosing the new
sample points randomly from the entire [0,10;0,10] range (thus satisfying eq~\ref{eq:ns_samples} exactly). The final set of sample 
points are shown by green crosses on Figure~\ref{fig:test_3min_3D}
and to construct the graph we have chosen $k=6$. To help visualise the
saddle point identification process, in the bottom panel of
Figure~\ref{fig:test_3min_3D} we show the state of the graph
just before it splits into two subgraphs corresponding to the two
larger basins.

We draw an energy landscape chart, shown in
the bottom left panel of Figure~\ref{fig:test_3min_3D}, in which the width
of the landscape at a given energy level is proportional to the phase space
volume enclosed by the subset of samples below that energy, as given by the nested sampling weights, $w_n$. Separate
basins are drawn according to our graph analysis.  Note that the ordering of the basins on the
  horizontal axis is arbitrarily chosen at each transition state, but
  their topological relationships are preserved. 
The gray shading in Figure~\ref{fig:test_3min_3D}
represents one standard deviation error in the overall phase space
volume. The error in the relative phase space volumes
of split basins is estimated as the standard deviation of the
multinomial distribution with generator probabilities equal to the
relative basin sizes. 

\begin{figure}[bthp]
\begin{center}
\includegraphics[width=6cm]{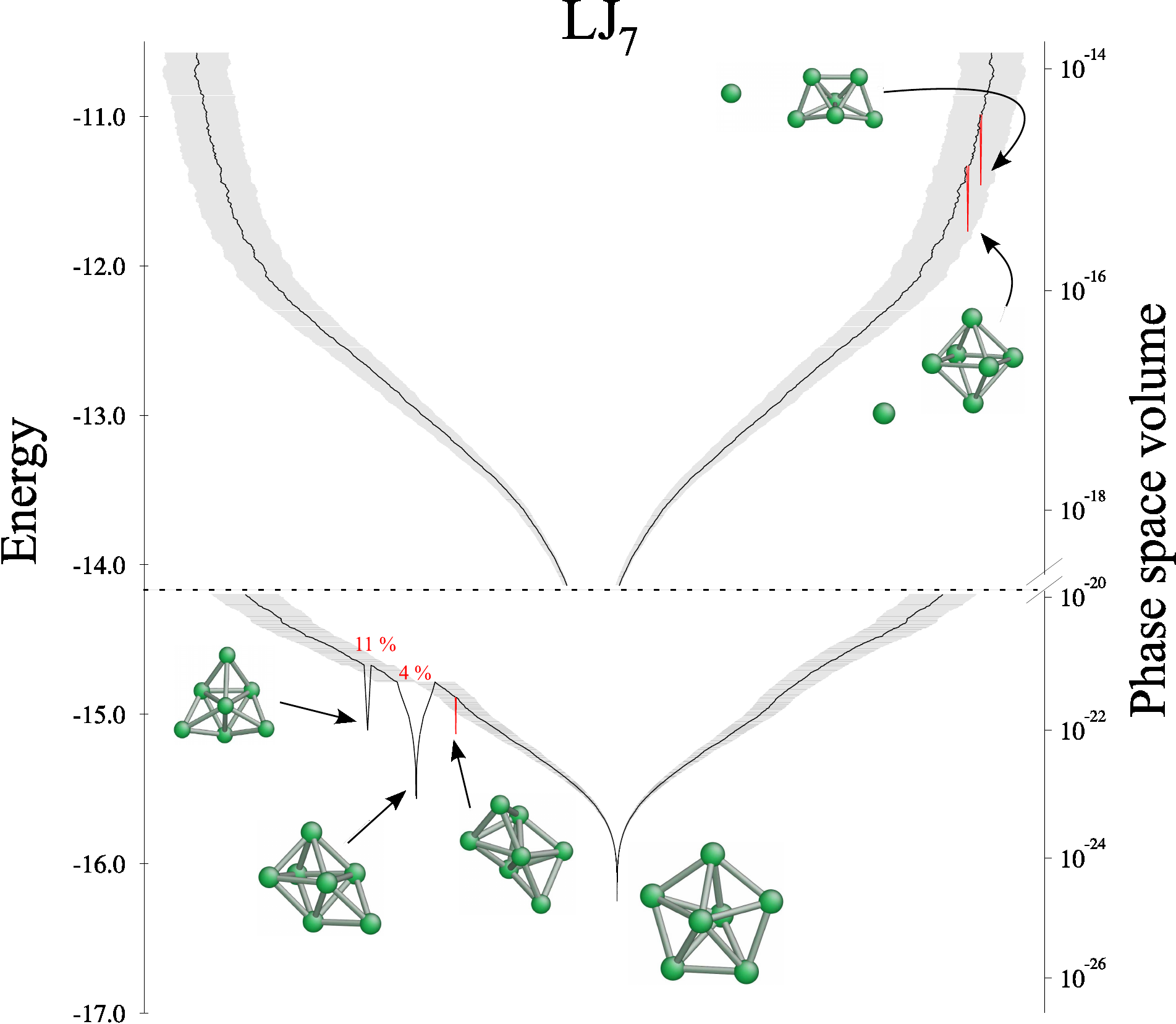}
\includegraphics[width=6cm]{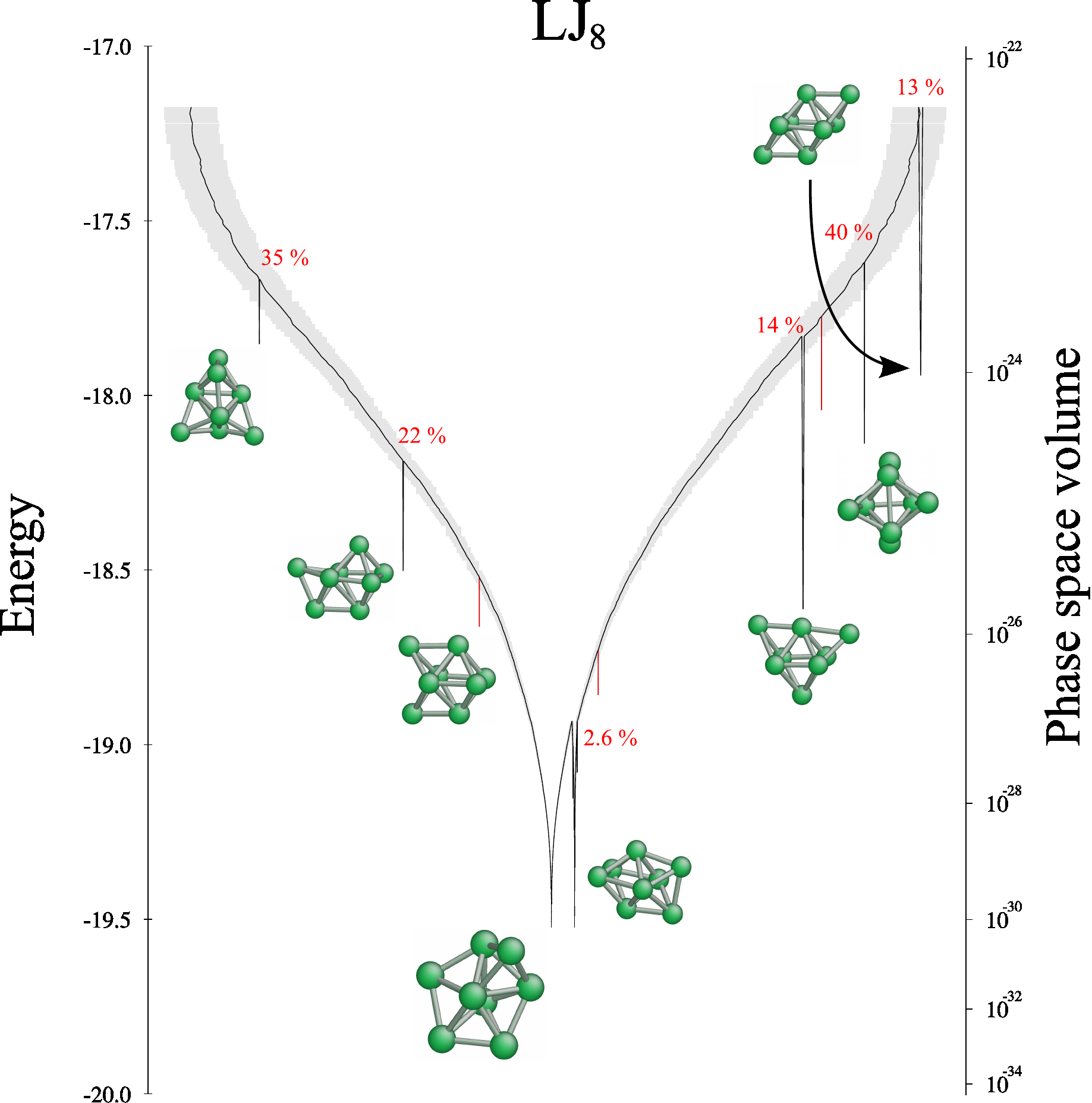}
\includegraphics[width=6cm]{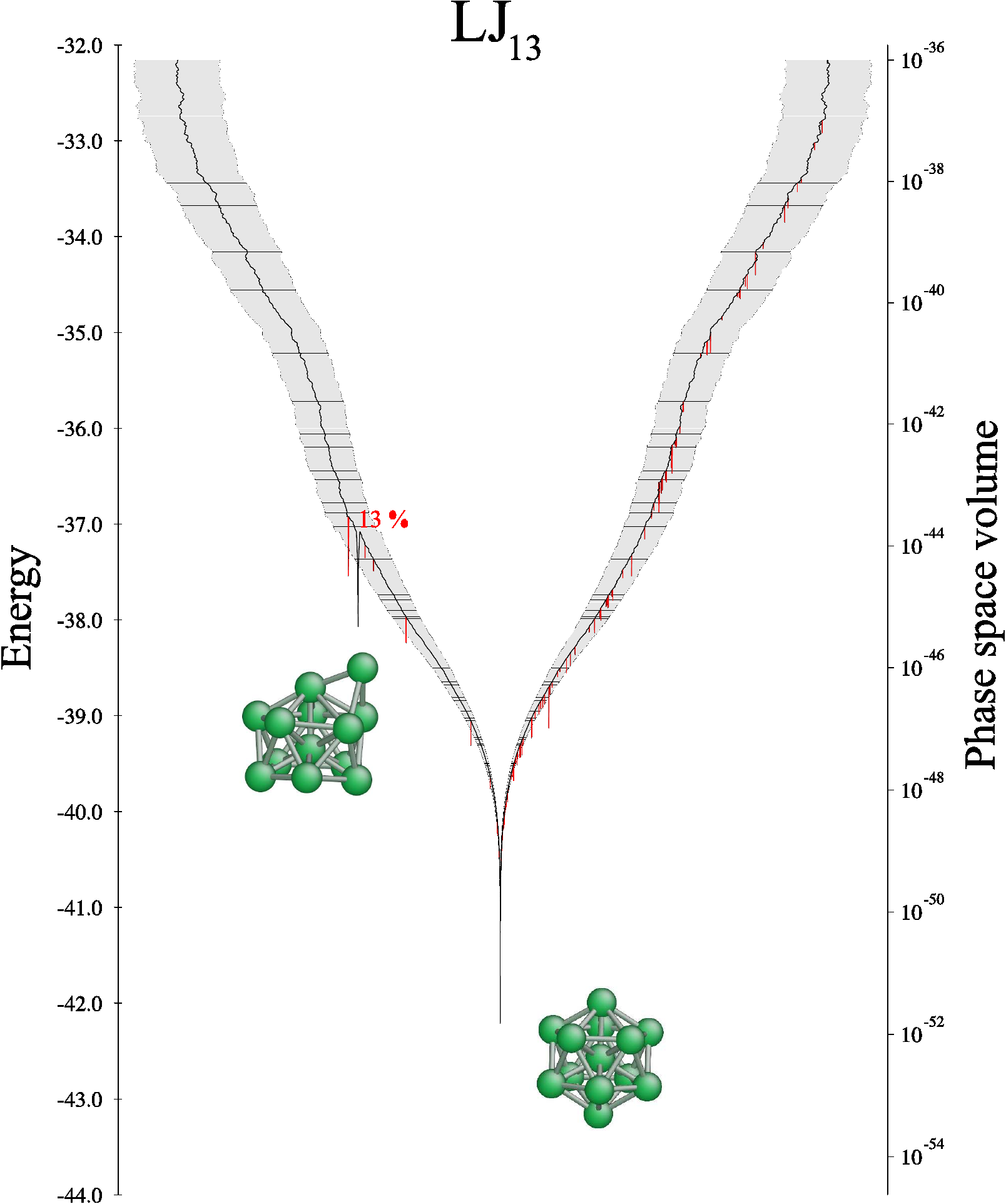}
\end{center}
\caption {Energy landscape charts of clusters of 7 (top left), 
  8 (top right), 13 (bottom left) and 36 (bottom right) Lennard-Jones atoms, (see Figure~\ref{fig:test_3min_3D} 
  for a detailed description of energy landscape charts).
   Here, basins where the error
  exceeds the basin size are coloured red.}
\label{fig:LJ_landscapes_small}
\end{figure}

\begin{figure}[bthp]
\begin{center}
\includegraphics[width=6cm]{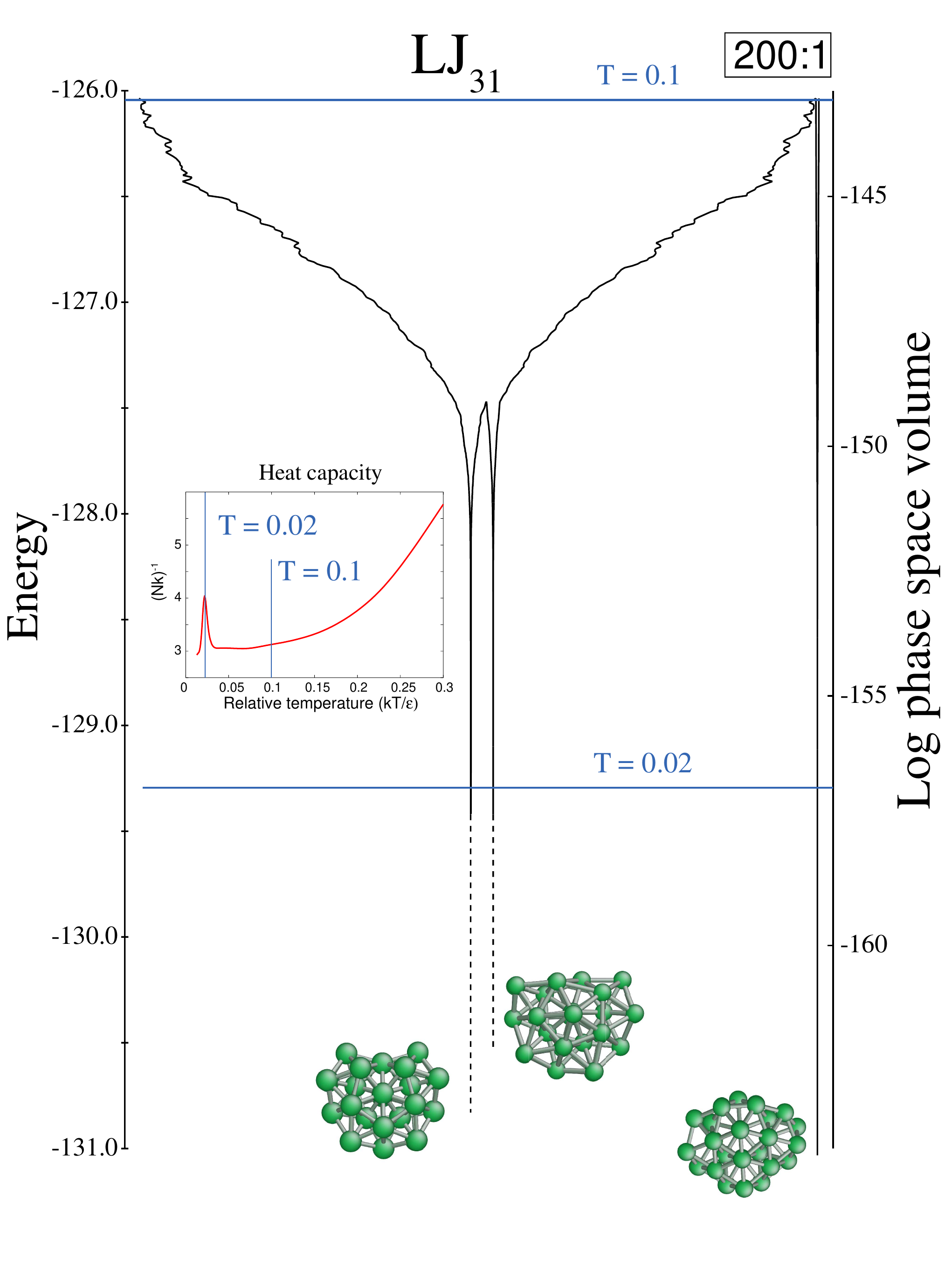}
\includegraphics[width=6cm]{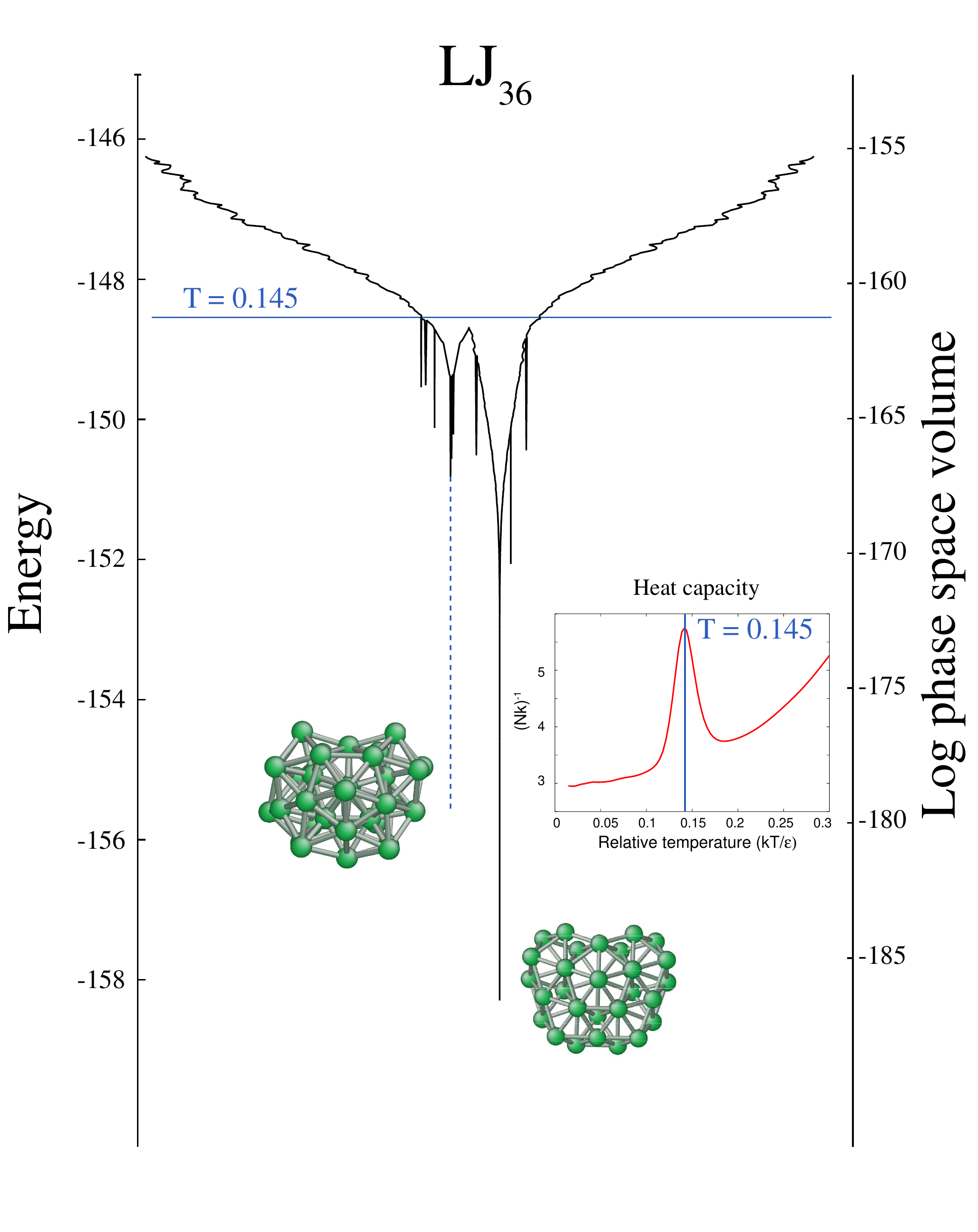}
\end{center}
\caption {Energy landscape charts LJ$_{31}$ (left)  and LJ$_{36}$ (right) clusters, (see Figure~\ref{fig:test_3min_3D} 
  for a detailed description of energy landscape charts).
   The insets show the heat capacity in the range of the first peaks. The expectation value of the energy (see eq~\ref{eq:energy}) corresponding to the specific temperatures marked are also shown on the
  energy landscape chart by blue lines. The boxed ratio for LJ$_{31}$ represents the phase space volume ratio of the basin containing the global minimum at its separation energy.}
\label{fig:LJ_landscapes_big}
\end{figure}

In order to construct the energy landscape charts for LJ clusters,
a distance metric between the configurations has to be constructed
that takes account of the exact symmetries of the Hamiltonian. The
metric that we use has been described elsewhere\cite{bib:so4}, it is
calculated in an auxiliary space in which configurations related by an
exact symmetry (translations, rotations and particle permutations) are
first mapped onto the same point by a continuous mapping. The
resulting energy landscape charts are shown in
Figure~\ref{fig:LJ_landscapes_small} for LJ$_7$, LJ$_8$,  LJ$_{13}$ and on Figure~\ref{fig:LJ_landscapes_big} for LJ$_{31}$ and LJ$_{36}$. Note that in this case and
in general for high dimensional systems, in contrast to the toy model, the
horizontal scale on which the phase space volume is represented has to
be an exponential function of the energy in order to fit the diagram
comfortably on the page.  It is particularly notable for LJ$_7$ that the two local minima with the
highest energies correspond to configurations in which one atom is in
the gas phase, and the others form LJ$_6$. Such a configuration is a
valid one for seven atoms in a box, and naturally appear in a nested
sampling run, because it samples the {\em entirety} of phase
space. Because one atom is in the gas phase, the phase space volume
associated with these local minima depends on the box size (in
contrast to the phase space volume of the local minima of the complete
cluster). For much larger boxes, the entropy of the gas atom would
dominate, as expected: matter sublimates at all temperatures in an
infinite perfect vacuum. Other such configurations are not shown because
they occur at higher energies. 

The energy landscape chart of
LJ$_{13}$  has a highly symmetrical global minimum. The
landscape has previously been mapped extensively and has at least 1478
local minima\cite{lj13}.  Our sample set was clearly too small to discover all of them, but this is not the
aim here. The figure shows an overall view of the PES, with its deep and
wide global minimum and no significant local minima at this resolution. In contrast, for smaller clusters, like LJ$_7$ or
LJ$_8$, narrow metastable states are already visible. The advantage of using nested sampling is that  we {\em do
  not} have to discover all local minima to be able to make qualitative statements
about the large scale features of the PES. Furthermore, the above difference between LJ$_{13}$ and
LJ$_7$ or LJ$_8$ cannot be gleaned from their respective disconnectivity graphs, even if they were mapped exactly. 
For larger or more complex systems, which have immense numbers of local minima, the nested sampling approach
will likely remain useful.

The smallest cluster for which a low temperature peak is present in the heat capacity is LJ$_{31}$, where the energy landscape chart shows that immediately above the peak (which is at $T\approx 0.02$) a handful of very similar states, distinct from the global minimum, dominate. At much higher temperature, $T=0.1$, at the separation point of the basin containing the global minimum, the system transitions between many distinct configurations, which is confirmed by MD simulations. 

The case of the LJ$_{36}$ cluster seems somewhat different in that near the energy value that corresponds to the temperature of the heat capacity peak, already above the separation point of the basin of the global minimum, a pronounced widening of the energy landscape can be observed, indicating that the number of available
configurations is very large. This behaviour has been well documented previously but for much larger clusters\cite{bib:lowT_transition}. The picture is confirmed by a short (10~ns) molecular dynamics run at $T=0.155$, in which we observed the system making transitions between many states with none of them dominating. 

\section{Free energy and a discrete order parameter}

A large part of solid state physics, chemistry and materials science
is concerned with determining phase diagrams. The existence of thermodynamic phase transitions can
be discovered using the appropriate response functions, e.g. as
illustrated above. However, the actual microscopic identification of the
different phases is much more subjective, since it requires
an externally defined {\em order parameter}, typically a
collective function of atomic coordinates.

The degree of arbitrariness in the choice of order parameter becomes a
major problem when dealing with phases that correspond to different
atomic structures, e.g. the various local minima of
clusters. Corresponding free energies can only be calculated once the
order parameter is defined, but in order to do that, one has to know
{\em in advance} what structures are to be distinguished---but in an
ideal world, that information should be the {\em result} of the free
energy calculation: the various phases correspond to the local minima
of the {\em free energy landscape}. Fluctuations at finite temperature make some ad-hoc order
parameters unusable, and degeneracies between equivalent structures
related by a permutation of atomic labels further complicates the task
of defining collective variables suitable to be used as order
parameters. 

Energy landscape charts suggest a different approach. 
Having explored the energy landscape at a given resolution,
we obtain a hierarchical tree of basins. The order
parameter that might correspond best to the natural philosopher's question
``Which state is the system in?'' is simply the identity of an energy
landscape basin or supra-basin (the latter can be formally defined as a collection
of basins each reachable from the others without having to traverse a
configuration with higher energy than the highest escape barrier from
the collection).  Accordingly, we label each basin and supra-basin,
and use this label as a {\em discrete order parameter}. Since every
sample point can then be assigned to a basin or supra-basin and
therefore to a particular value of this order parameter, computing
free energies by using partial partition function (summing over just the samples in a given basin) is straightforward. 

We use this approach to determine the free energy difference between
the icosahedral metastable minimum and the global minimum of the famous LJ$_{38}$ cluster.\cite{bib:lj38_1} Previous estimates of the relative sizes of the corresponding supra-basins range
from 20:1\cite{bib:wales_scale_disconnectivity} (based on the number of local
minima found in each supra-basin below the lowest transition point) to 10000:1\cite{ChrisPickard_private_communication} (based on the relative frequency of finding the two minima using random search). 

\begin{figure}[bthp]
\begin{center}
\includegraphics[width=7cm]{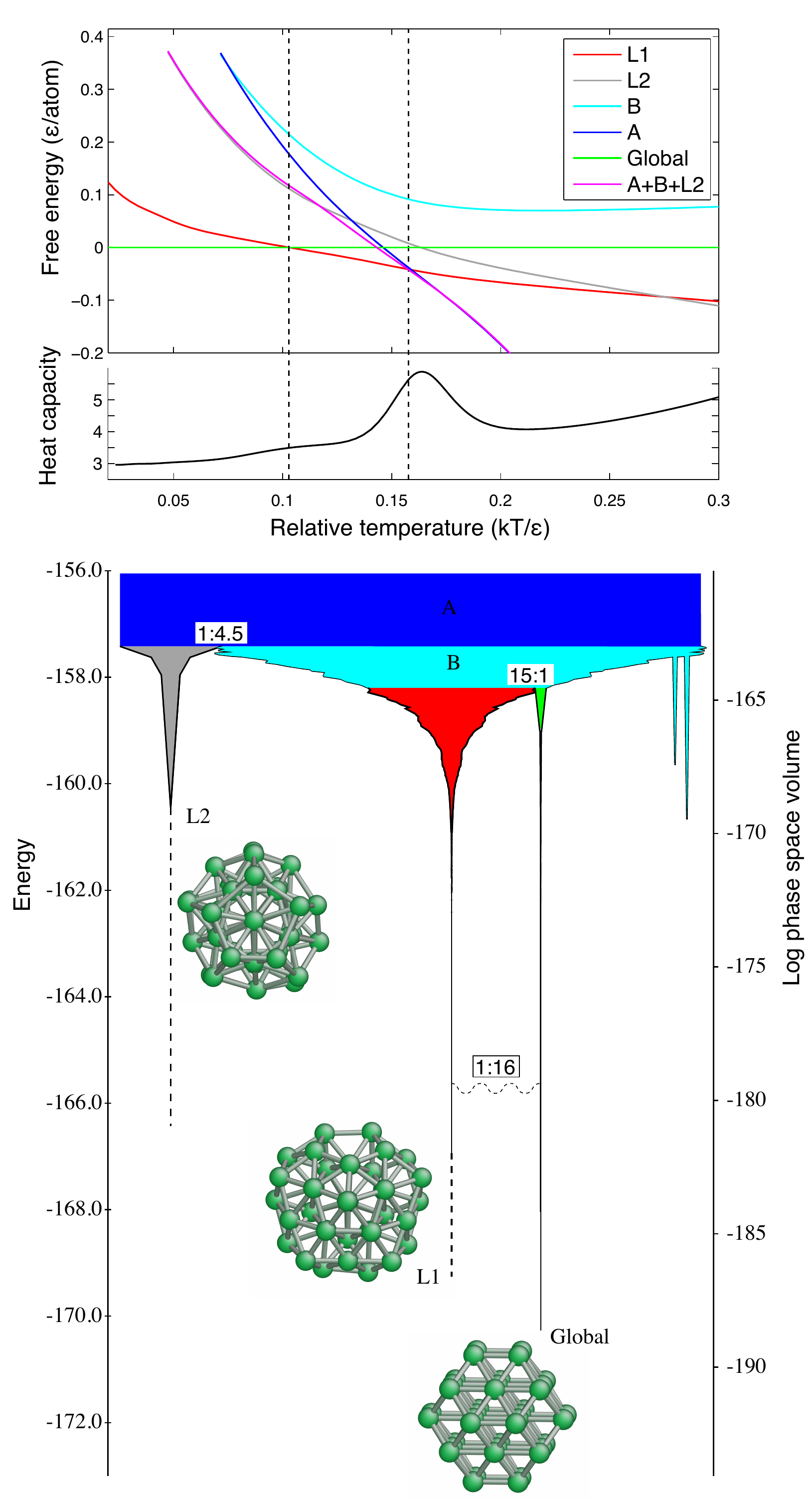}
\end{center}
\caption {Free energy of basins (top) and energy landscape chart (bottom) of LJ$_{38}$. The coloured regions identify the major basins: global fcc minimum (green), icosahedral local minimum (L1, red), alternative icosahedral structure (L2, gray). The dashed lines indicate the depth of the basins as obtained with direct minimisation starting from the lowest local sample of nested sampling. The wavy line between L1 and the global minimum on the energy landscape chart indicates the approximate energy of the minimum energy path connecting the to minima\cite{bib:walesLJ38freeenergy}. The relative phase space volumes of the basins at the separation point are shown by the boxed ratios. The light blue region represents states in which the fcc and lowest energy icosahedral states cannot be distinguished, the dark blue region includes states in which L2 states also become indistinguishable. The top panel shows the free energy associated with each coloured region. For reference, the heat capacity curve is also plotted again. }
\label{fig:LJ38_freeenergy}
\end{figure}

We carried out the nested sampling calculation for this analysis using $K=64000$ for LJ$_{38}$. The resulting energy landscape chart is shown on~Figure~\ref{fig:LJ38_freeenergy}. It shows three distinct large basins corresponding to the global minimum and two icosahedral minima, which we label "Global", "L1" and "L2" (we have ignored two apparent basins that are very small and not stable as the parameters of the graph analysis algorithm are varied). As the energy increases from the global minimum, first the global minimum and L1 merge (these states are labelled "B"), then this suprabasin merges with L2 (these states are labelled "A"). We plot the relative free energies of the various regions on the top panel, showing that the L1 state becomes the stable at about $T\approx0.1$ and the A state becomes stable at $T\approx0.16$. These transitions show very good agreement with the positions of the peaks in the heat capacity curve (also shown). The match is not exact because the basin identification is based on energy, whereas the natural variable of the heat capacity is temperature, and these only have a one-to-one correspondence in the thermodynamic limit. The energy landscape chart has therefore automatically identified the identity of the various distinct thermodynamically stable states without external input such as an order parameter. 

The relative sizes of the various basins at their separation energies are also indicated. In particular, the phase space volume ratio of the lowest energy icosahedral minimum to that of the global minimum is about 15:1. It is interesting to note that the separation energy of these basins (basically the energy at which the structures become identifiable as distinct) is much higher than the highest point on the minimum energy path\cite{bib:walesLJ38freeenergy}. The latter is indicated on~Figure~\ref{fig:LJ38_freeenergy} by a wavy line, and the phase space volume ratio at that energy is 1:16, i.e. the basin of the global minimum is much larger there. 


\section{Conclusion}

We described a new framework for efficiently sampling complex energy
landscapes, based on nested sampling. This ``top-down'' approach is
inherently unbiased and its resolution can be adjusted to suit the
available computational resources. Although it can be used as a tool
to search for minima, we expect that
its main strength will be that it can provide an approximate 
picture of the large scale features of the landscape
using only modest resources. Beyond this qualitative description, the
sample points can be used to evaluate the partition function with good 
accuracy at arbitrary temperatures, and hence also the expectation values of thermodynamic
observables, such as response functions. 

Furthermore, the topological analysis of the samples can be used to
discover large scale basins
and identify them with the macroscopic states of the system. We defined
an order parameter which
indexes the basins. The knowledge of the phase space volumes
associated with each such basin allows the direct evaluation of the
free energy corresponding to each value of this order parameter, and
hence give information on the relative stability of the macroscopic
states, without need for {\em a priori} definitions of these states. 

We demonstrated nested sampling in the well studied system of Lennard-Jones clusters,
where the efficiency of evaluating the heat capacity was more than an order of magnitude better than
that of parallel tempering, without using any prior knowledge of the location of the global minima. Because the efficiency gain comes from the natural handling of first order phase transitions,
we expect even better results in bulk systems, whose study is already underway.
\\
{\textbf{Acknowledgement}\\
  The authors are indebted to John Skilling, Daan Frenkel and David
  Wales for carefully reading the manuscript and to Ben Hourahine,
  Noam Bernstein, Mike Hobson, Farhan Feroz and Mike Payne for
  extensive discussions.  The work has been partly performed under the
  Project HPC-EUROPA (RII3-CT-2003-506079), with the support of the
  European Community - Research Infrastructure Action under the FP6
  ``Structuring the European Research Area'' Programme. LBP acknowledges
  support from the E\"otv\"os Fellowship of the Hungarian State and
  the hospitality of the Engineering Laboratory in Cambridge. GC would
  like to acknowledge support from the EPSRC under grant number
  EP/C52392X/1. 
}

\bibliography{NS1_ref}
\bibliographystyle{h-physrev3}

\end{document}